\documentclass[lettersize,journal]{IEEEtran}
\usepackage{amsmath,amsfonts}
\usepackage{algorithmic}
\usepackage{algorithm}
\usepackage{array}
\usepackage[caption=false,font=normalsize,labelfont=sf,textfont=sf]{subfig}
\usepackage{textcomp}
\usepackage{stfloats}
\usepackage{url}
\usepackage{verbatim}
\usepackage{graphicx}
\usepackage{cite}
\usepackage{multirow}
\usepackage{booktabs}
\usepackage{xcolor}
\begin{document}

\title{Agentic UE-CoMIMO for 6G Terminals: From Virtual Antenna Augmentation to AI-Native Virtualization}

\author{Chao-Kai~Wen,~\IEEEmembership{Fellow,~IEEE}, Yen-Cheng~Chan, Lung-Sheng~Tsai, Pei-Kai~Liao,~\IEEEmembership{Senior~Member,~IEEE}, and~Geoffrey~Ye~Li,~\IEEEmembership{Fellow,~IEEE}

\thanks{{C.-K.~Wen} is with the Institute of Communications Engineering, National Sun Yat-sen University, Kaohsiung 80424, Taiwan (e-mail: {\rm chaokai.wen@mail.nsysu.edu.tw}).}

\thanks{{Y.-C. Chan} is with the Institute of Electrical
Control Engineering, National Yang Ming Chiao Tung University, Hsinchu
300, Taiwan (e-mail: {\rm ycchan7250@gmail.com}).}

\thanks{{L.-S.~Tsai} and {P.-K.~Liao} are with MediaTek Inc., Hsinchu, Taiwan (e-mail: {\rm Longson.Tsai@mediatek.com, pk.liao@mediatek.com}).}

\thanks{{G.~Y.~Li} is with the Department of Electrical and Electronic Engineering, Imperial College London, SW7 2AZ London, U.K. (e-mail: {\rm geoffrey.li@imperial.ac.uk}).}

}

\markboth{IEEE Communications Magazine}%
{Wen \MakeLowercase{\textit{et al.}}: Agentic UE-CoMIMO for 6G: From Virtual Antenna Augmentation to AI-Native Virtualized Terminals}


\maketitle

\begin{abstract}
End-user-centric collaborative MIMO (UE-CoMIMO) lets nearby devices form a virtual multi-antenna terminal to overcome the antenna limitations of individual user equipment. Extending such cooperation to communication, sensing, computing, and task-relevant information exchange requires a control layer that can interpret user intent, select cooperation mechanisms, and replan as conditions change. This article introduces Agentic UE-CoMIMO, in which device micro-agents, a smartphone or CPE hub agent, and edge/network agents coordinate device participation, relay modes, traffic splitting and duplication, compute placement, semantic-token exchange, and topology reconfiguration. Two system-level scenario studies on creator-centric live streaming and wearable-collaborative blind-spot sensing compare the proposed controller with capability-matched adaptive baselines. The results show that, by anticipating changes and preparing cooperation and fallback actions in advance, agentic control sustains high-quality streaming for longer and maintains blind-spot warnings through device outages. We also discuss the associated standardization, interoperability, trust, and validation challenges.
\end{abstract}


\section{Introduction}
The evolution of MIMO toward 6G cannot rely on adding more antennas at the base station alone. Massive MIMO, multi-TRP transmission, and extremely large-scale arrays continue to increase the spatial capability of the network side \cite{Lu2024XLMIMO}. The user-equipment (UE) side, however, follows a different scaling law because form factor, antenna spacing, power, and thermal budget limit the number of antennas that a terminal can support. Lightweight wearables may have only one or two antennas, and even smartphones support only a few in sub-6 GHz bands. The UE side therefore cannot support additional spatial streams even when the network can offer them.

End-user-centric collaborative MIMO (UE-CoMIMO) addresses this bottleneck by allowing a primary UE (Pr-UE) to recruit nearby collaborative UEs (Co-UEs) and local infrastructure into a virtual antenna system \cite{Tsai2024ComMag}. Its gains in effective channel dimensionality and throughput have been demonstrated analytically and experimentally using frequency-translation relay structures \cite{Wen2025TWC} and relay-assisted carrier aggregation \cite{Chen2025RACA}. These mechanisms provide the physical basis for user-side cooperation.

The next question is how these mechanisms should be used when several devices and tasks coexist. A user may carry AI glasses, a smartphone, and a smartwatch while interacting with customer-premises equipment (CPE), vehicles, or edge servers. The same cluster may need to sustain a first-person live stream, monitor a blind spot for hazards, and host an AI assistant \cite{Jiang2026IntentionAware}. The appropriate devices, roles, relay modes, and compute placement therefore depend on user intent and device state as well as channel quality. This article introduces \emph{Agentic UE-CoMIMO}, an extension of UE-CoMIMO in which a user-side control layer interprets intent, decomposes the task into communication, sensing, and computing objectives, plans the required cooperation, selects the available mechanisms, maintains task state, and replans according to observed outcomes.

The term \emph{agentic} is used here to distinguish this control process from conventional context-aware cross-layer adaptation. A conventional controller typically optimizes a predefined objective over a fixed set of variables and actions. In Agentic UE-CoMIMO, the objective, participating devices, and required mechanisms may change as the task proceeds. Consistent with the broader agentic-AI literature, where planning, memory, reflection, and tool use are identified as key capabilities \cite{Liang2026LLM4COM}, the proposed control layer combines intent interpretation, task decomposition, mechanism selection, state and memory, outcome monitoring, and feedback-driven replanning. Conventional optimization or learned policies can still operate within a selected mechanism, for example for resource allocation or relay configuration. The agentic layer determines which mechanisms should be activated and when they should be changed. 

Agentic UE-CoMIMO is related to several existing research directions but addresses a different control scope. Recent agentic-LLM and multi-agent wireless-control studies consider network-side resource management, network slicing, and multi-AP coordination, where agents reason over network states, plan actions, and invoke wireless control mechanisms \cite{Liang2026LLM4COM,Li2026ComAgent,Tong2025WirelessAgent,Zheng2026CrossModule,Zhang2025MARL6G}. These studies primarily focus on network- or AP-side control rather than the coordination of a heterogeneous end-user device cluster. Multi-connectivity combines multiple links for a UE, while multi-device orchestration coordinates resources across devices. Cooperative sensing focuses on distributed sensing and information fusion, edge intelligence \cite{Zhou2019EdgeIntelligence} addresses compute placement and collaborative inference, and intent-based networking typically translates service or operator intent into network configuration. Agentic UE-CoMIMO instead places the control plane on the user side and jointly manages device participation, physical-layer cooperation, sensing roles, compute placement, and semantic exchange under antenna, battery, thermal, trust, and radio-resource constraints.

\section{Agentic UE-CoMIMO: Architecture for AI-Native Virtualized Terminals}
\label{sec:architecture}

This section presents the architecture of Agentic UE-CoMIMO. We first explain how it extends UE-CoMIMO from physical-layer antenna cooperation to coordinated communication, sensing, computation, and semantic exchange. We then describe the roles of device micro-agents, smartphone/CPE hub agents, and edge/network agents.

\subsection{From UE-CoMIMO to Agentic UE-CoMIMO}

UE-CoMIMO, also called terminal-cooperative MIMO in the broader beyond-5G/6G context, treats nearby user-side devices as a cooperative virtual antenna system \cite{Tsai2024ComMag,XGMF2025TerminalCooperativeMIMO}. These devices may belong to different trust domains. A personal cluster of glasses, a smartphone, and a smartwatch is relatively straightforward to authorize, whereas indoor CPEs and access points or outdoor vehicular and roadside nodes may provide more stable and better-positioned assistance but raise additional authorization and incentive issues. UE-CoMIMO is therefore a framework for forming virtualized spatial resources from available nearby devices rather than a single device pair.

The initial UE-CoMIMO vision focused on physical-layer augmentation. A Co-UE can provide a diversity branch, help form a higher-rank effective MIMO channel, or extend the virtual array for more accurate localization \cite{Tsai2024ComMag}. Another design choice is the relay layer. Latency-critical, high-rate services favor L1 frequency-translation relays, which forward signals at the physical layer without packet decoding or buffering. Less time-sensitive services can use L2/L3 relays, which decode and forward packets and therefore support packet-level processing and cross-RAT routing.

Agentic UE-CoMIMO extends virtual antenna aggregation with four components: agentic control, distributed computing, semantic-token exchange, and integrated communication, sensing, and perception. It retains the physical-layer mechanisms and adds an intelligence layer for their coordination. UE-CoMIMO defines the available mechanisms, while Agentic UE-CoMIMO determines how they are used according to the current task. The control objective therefore includes not only spectral efficiency and rank, but also quality of experience, sensing confidence, energy consumption, and thermal state.

\subsection{Hierarchical Agentic AI}

\begin{figure}[t]
    \centering
    \includegraphics[width=\linewidth]{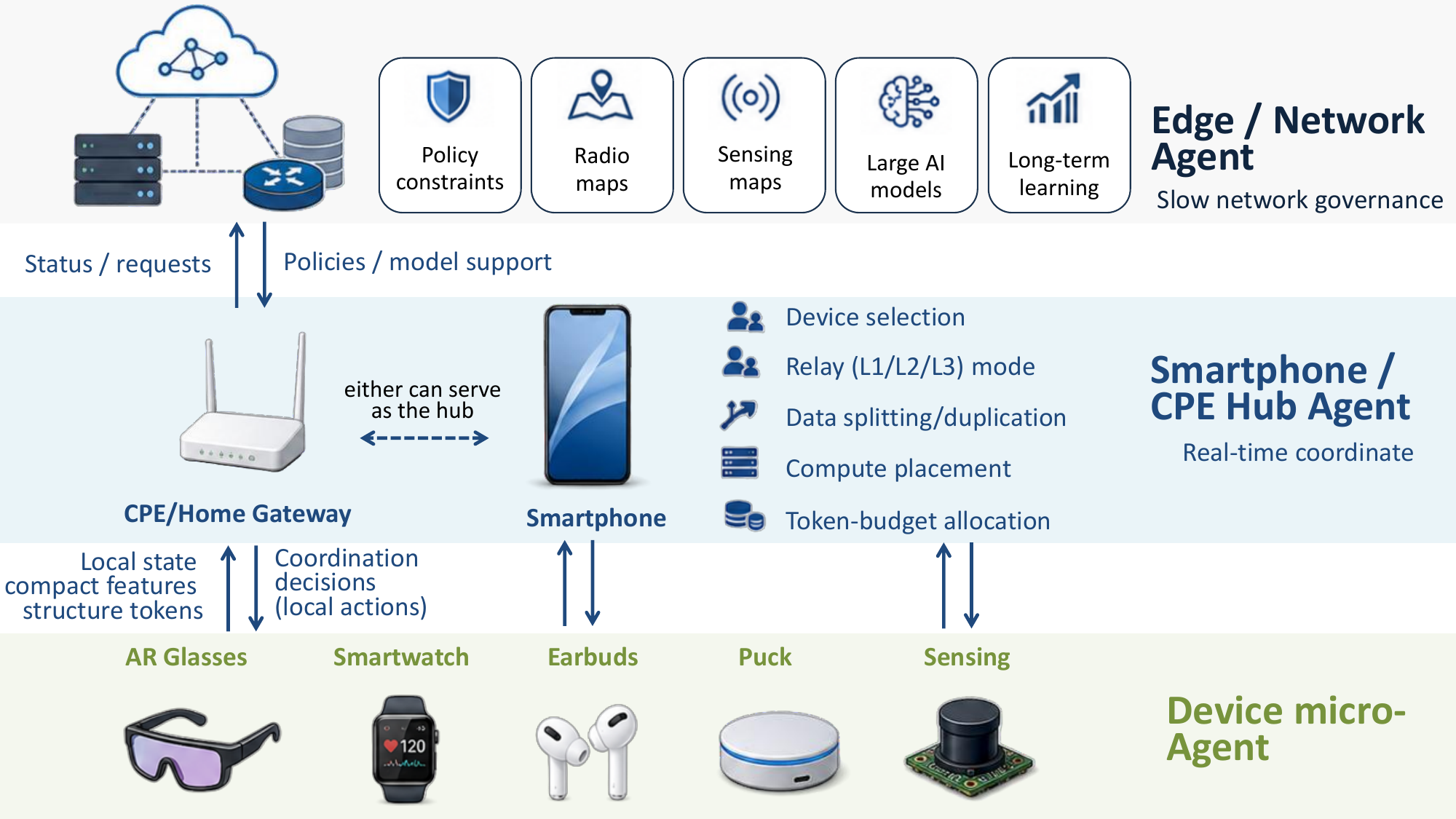}
    \caption{Hierarchical Agentic UE-CoMIMO architecture. Lightweight devices host micro-agents, a smartphone or CPE acts as the hub agent for real-time coordination, and the edge/network agent supplies policy constraints, radio and sensing maps, and large AI models.}
    \label{fig:architecture}
\end{figure}

A practical agentic architecture distributes intelligence across different locations. Lightweight devices are constrained by power, thermal budget, and computing capability and are therefore unsuitable for complex decision making. The edge network provides greater resources and broader visibility but may not respond quickly enough to local events such as blockage, mobility, or overheating. A smartphone or CPE provides an intermediate coordination point. As illustrated in Fig.~\ref{fig:architecture}, we therefore use three agent layers: a device micro-agent, a smartphone/CPE hub agent, and an edge/network agent. This structure separates fast local control from slower network-level coordination.

\subsubsection{Device Micro-Agent}

Micro-agents reside on lightweight devices. Their role is to provide timely local awareness rather than global optimization. A micro-agent reports locally observable states, such as link quality, battery level, and temperature. It also tags the traffic it generates by type and deadline. When computing resources allow, it can replace raw samples with lightweight features or structured tokens. If local conditions become critical, it sends a fallback request to the hub agent. 

\subsubsection{Smartphone/CPE Hub Agent}

The hub agent is the real-time coordinator. A smartphone is a natural hub in mobile personal-area scenarios because of its computing capability, multiple radio interfaces, and larger battery. Indoors, a CPE or Wi-Fi AP can take this role because of its stable power supply, large aperture, and favorable placement. The hub decides which devices participate and in what roles, which relay or traffic mode carries the data, where computation runs along the device-edge-cloud path \cite{Zhou2019EdgeIntelligence}, and how the semantic-token budget is allocated. It serves as the control plane of the virtualized terminal. It need not process all payload data but determines which mechanisms are active for the current task.

\subsubsection{Edge/Network Agent}

The edge/network agent operates at the edge server, RAN controller, or network AI function. It provides information and capabilities that a local cluster cannot readily generate, including radio and sensing maps, inter-user coordination, large-model inference, and a \emph{policy envelope}. This envelope defines the parameter ranges, resource budgets, and security policies within which the cluster may operate. Agentic UE-CoMIMO therefore separates \emph{slow network-level coordination from fast local agentic control}. The network specifies what is allowed, while the hub selects actions within those limits, as described in Section~\ref{sec:control_and_semantic}.

\section{Agentic Control and Semantic Intelligence}
\label{sec:control_and_semantic}

\begin{figure}[t]
    \centering
    \includegraphics[width=\linewidth]{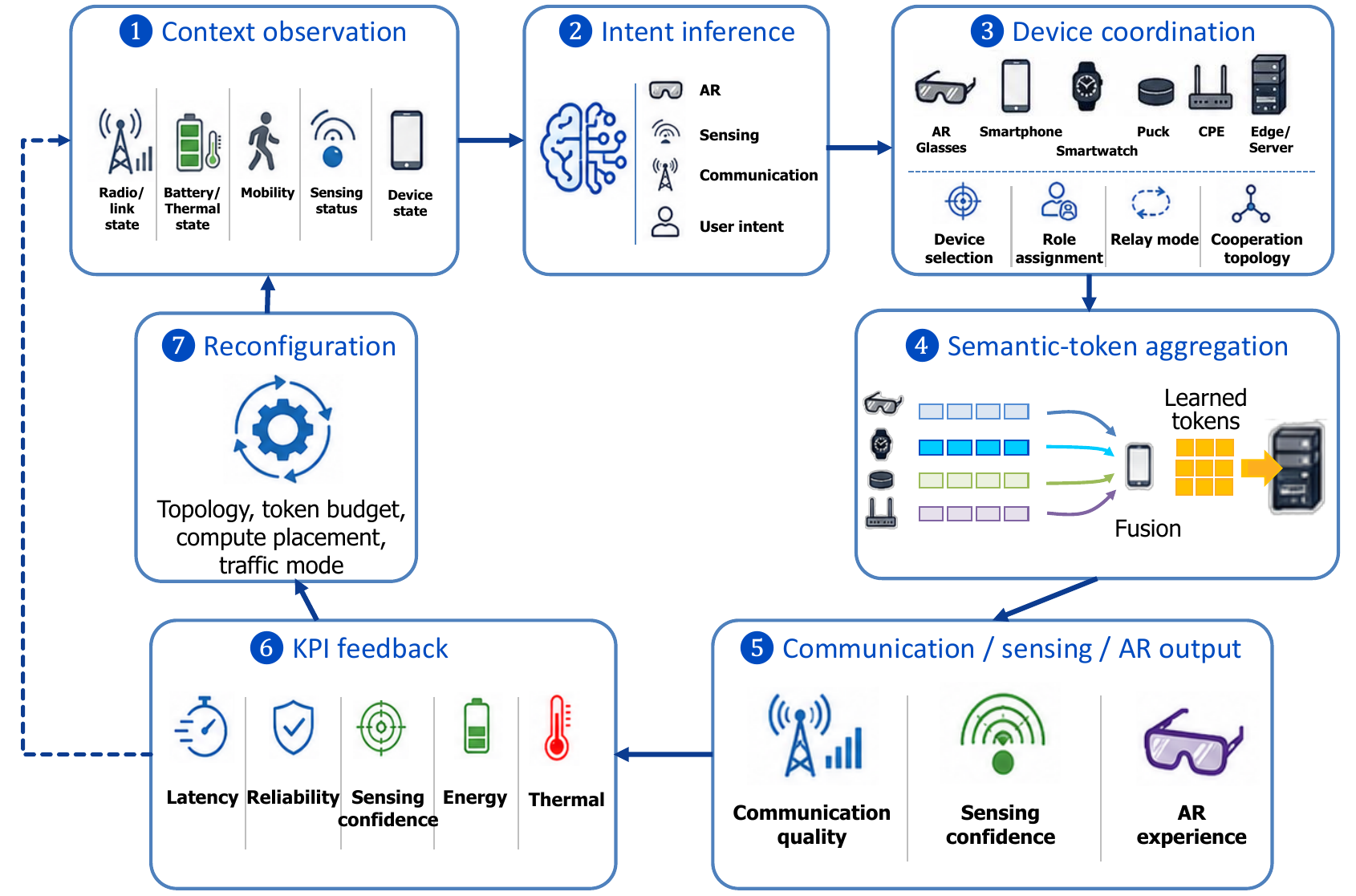}
    \caption{Agentic control and semantic intelligence loop. Context observation and intent inference guide device coordination and token aggregation. The resulting task quality is fed back to reconfigure the topology, token budget, compute placement, and traffic mode.}
    \label{fig:control_loop}
\end{figure}

As illustrated in Fig.~\ref{fig:control_loop}, cooperation is treated as a closed-loop control process. The system observes the context, interprets the current intent, selects a cooperative topology and mechanisms, exchanges compact semantic information, monitors the resulting task quality, and revises its decisions accordingly.

Here, \emph{agentic} refers to a controller that can modify how the control problem is organized as the task evolves. A conventional context-aware cross-layer controller typically adapts variables within a predefined objective and action structure. The agentic layer can additionally interpret intent, decompose the task, select and combine heterogeneous mechanisms, maintain task state, monitor outcomes, and revise earlier decisions when the objective or operating conditions change. Closed-form optimization or learned policies can still be used within each selected mechanism.

Three capabilities are emphasized in the evaluations in Section~\ref{sec:two_use_cases}. \emph{Prediction} considers the expected evolution of the system rather than only the current state. \emph{Intent interpretation} maps the service objective to the quantities that the controller trades off, so the same cluster can behave differently under throughput, battery-life, or safety objectives. \emph{Feedback-driven replanning} uses device-state reports and downstream task quality to revise earlier decisions, for example when a recruited device becomes unavailable. 
Task decomposition, mechanism selection, and memory support these functions and are held fixed in the comparisons. 

These functions do not necessarily require a large AI model. In Section~\ref{sec:two_use_cases}, we implement them using interpretable rules and state machines so that the measured gains reflect the control functions rather than the capabilities of any particular AI model. Richer observations or a larger action space alone are likewise insufficient to establish an agentic gain. The evaluations therefore use capability-matched baselines with matched observations and action spaces, but without forward planning, intent interpretation, or feedback-driven replanning.

\subsection{Intent-Aware Multi-Device Coordination}

Consider the candidate devices around a user, such as glasses, a smartphone, a smartwatch, a CPE or access point, or a vehicle. At each decision instant, the hub agent chooses a cooperating subset, assigns each device a role, and sets the relay and traffic mode, compute placement, and token budget. Personal devices can be recruited directly, whereas shared, vehicular, or public devices are subject to authorization and incentive constraints.

The selection depends on the current service objective. The hub weighs communication, sensing, and perception quality against energy consumption, coordination overhead, and thermal stress. The relative importance of these quantities depends on the current intent. A throughput-oriented upload, a latency-critical interaction, and a safety-critical sensing task may therefore lead the same cluster to different configurations, and the intent may change during a session. The control objective is consequently broader than spectral efficiency alone.

\subsection{Semantic-Token Aggregation}

Exchanging raw observations is costly and often unnecessary. Raw CSI, sensing echoes, and video frames may exceed what a lightweight device can afford to transmit and may reveal more information than the task requires. Agentic UE-CoMIMO therefore exchanges compact, task-aware semantic tokens that may carry, for example, a dominant channel or sensing feature, a target-confidence value, or a blockage indicator.

A semantic token consists of a task-relevant payload, represented by structured features or an embedding, together with metadata for inter-device interpretation. The metadata identifies the task and source and may include uncertainty or confidence, freshness or timestamp, trust or reliability, and representation resolution. These fields allow a receiving agent to determine how the token should be used, weighted, updated, or discarded. This interoperability abstraction distinguishes semantic-token exchange from conventional feature exchange, in which the feature representation and its use are usually predefined within a specific inference pipeline.

This concept is related to conventional semantic communication, which focuses on conveying task-relevant information over a transmitter-receiver link \cite{Zhang2023Semantic}. In Agentic UE-CoMIMO, tokens instead provide a \emph{multi-device cooperation interface} among devices with different capabilities. A resource-constrained device may generate tokens using thresholding or CFAR-like processing, whereas a more capable device may use a learned encoder. The accompanying metadata allows the hub to interpret and fuse these heterogeneous outputs. The blind-spot study in Section~\ref{sec:two_use_cases} directly uses device reliability, target confidence, and a device-failure indication to support fusion and replanning.

\subsection{Task-Driven Topology Reconfiguration}

A key distinction from conventional cooperative MIMO is that the cooperating set need not remain fixed after transmission or sensing begins. Downstream task quality is fed back to the controller and can trigger changes in the topology itself. The hub may recruit another device or request higher-resolution tokens when confidence drops, duplicate critical packets when latency requirements become stricter, or move computation away from a device that is overheating. The two studies in Section~\ref{sec:two_use_cases} illustrate this closed loop for communication and sensing tasks.

\begin{figure}[t]
    \centering
    \includegraphics[width=\linewidth]{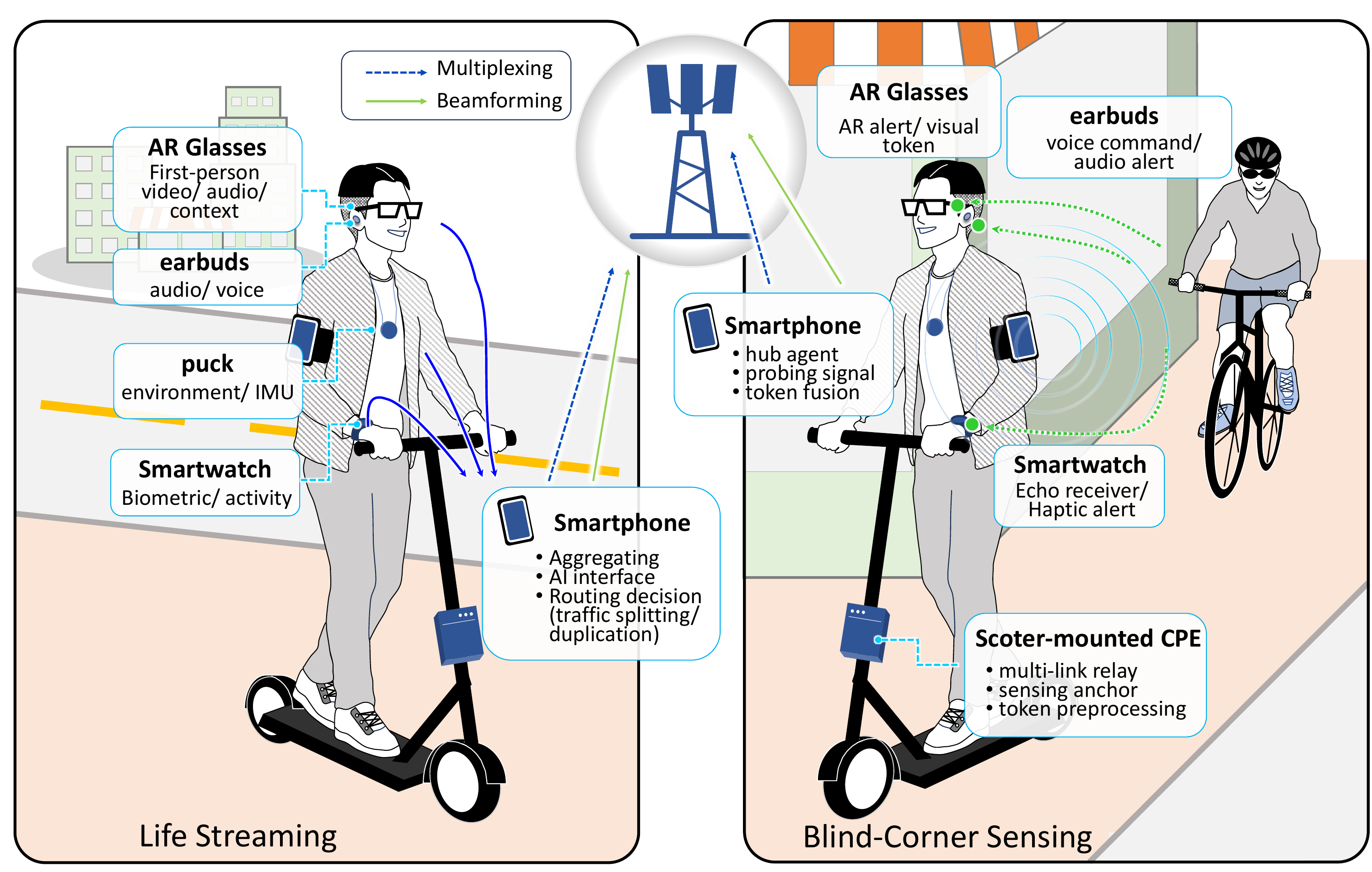}
    \caption{Representative creator-centric use cases: multi-device collaborative live streaming and wearable-collaborative blind-spot sensing. The same cluster of glasses, smartphone, smartwatch, puck, and scooter-mounted CPE is reconfigured to support first-person uplink, sensing-token exchange, and safety warning.}
    \label{fig:use_cases}
\end{figure}

\section{Two Representative Use Cases}
\label{sec:two_use_cases}

In this section, we present two representative use cases: creator-centric live streaming with CPE assistance and wearable-collaborative blind-spot sensing and warning.

\subsection{Creator-Centric Live Streaming with CPE Assistance} 

\paragraph{A General Description}
In creator-centric live streaming, a creator continuously uploads first-person video and audio through lightweight AI glasses, either outdoors while traveling or riding, or indoors in a studio, mall, or exhibition hall. Both settings require a stable, low-latency uplink over a long session, although the main bottlenecks differ. Outdoors, cell-edge coverage and body blockage dominate, whereas indoors, penetration loss and congestion become more important. In either case, lightweight glasses alone may not have sufficient antenna, battery, and thermal resources to sustain the stream. 

As depicted in Fig.~\ref{fig:use_cases}, the hub agent coordinates the creator's nearby devices as a virtualized terminal. It can recruit a smartphone or puck to assist the glasses, split bulk video across links, duplicate critical packets, and move transmission or computation away from the glasses as their temperature rises. Indoors, a CPE or Wi-Fi AP is an attractive Co-UE because of its stable power, larger aperture, and favorable placement. It can serve as a relay, virtual receiver, or local computing host \cite{Jiang2026IntentionAware}. The control objective is the creator's quality of experience over the entire session, including stable visual quality, low tail latency, energy consumption, and stream continuity across the outdoor-to-indoor transition.

\paragraph{An Illustrative Scenario Study}

We evaluate the agentic layer through a system-level simulation of this scenario. A user wearing AI glasses streams first-person video while walking from an outdoor street into a venue and then remaining indoors for an extended period. The glasses have no low-band antenna and limited thermal headroom, so they offload video over 5~GHz Wi-Fi to a smartphone hub agent. The smartphone maintains the cellular uplink and can transmit directly, relay over 3.5~GHz or 700~MHz, split packets with a nearby shared CPE, or delegate the uplink entirely to that CPE and turn off its own cellular radio. Using scene information from the glasses, the hub anticipates the outdoor-to-indoor (O2I) transition and selects the transmission mode accordingly. Its objective is to avoid stalls, control the temperature of the glasses, and extend battery life rather than maximize throughput. The complete geometry, channel, device, and control models, together with all parameters and their sources, are provided in the supplementary material \cite{WenSupplement2026}.

The comparison is designed to isolate the source of the performance gain. In addition to fixed and reactive cellular baselines and an ablation without the predictive governor, we include a \emph{capability-matched myopic controller}. This non-agentic controller observes the same variables as the proposed policy, including link rates, CPE load, temperature, buffer state, battery level, and current scene complexity. It also has access to the same action space, but greedily maximizes the current-step QoE without prediction, intent interpretation, or multi-step planning.

\begin{figure}[t]
    \centering
    \includegraphics[width=\linewidth]{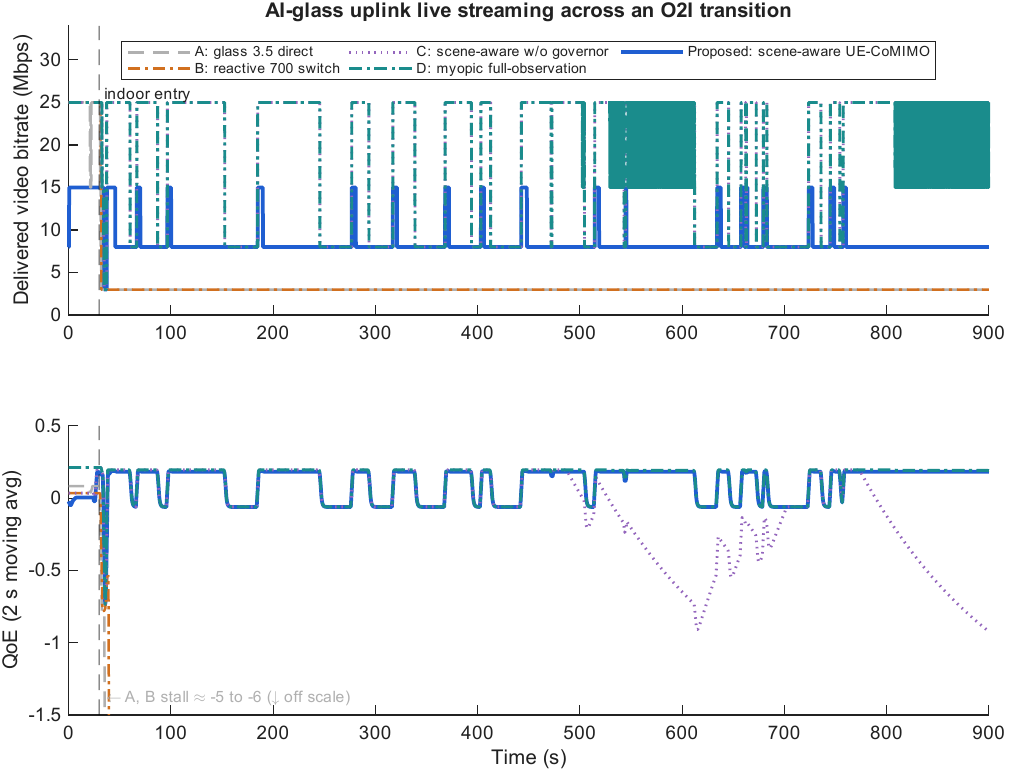}
    \caption{System-level evaluation in the creator-centric live-streaming scenario. The top panel shows the delivered video bitrate, and the bottom panel shows the smoothed QoE for the fixed (A) and reactive (B) cellular baselines, the ablation without the predictive governor (C), the capability-matched myopic controller (D), and the proposed agentic policy. The venue is entered at 30~s, and the indoor dwell lies inside an RF dead zone where A and B stall. Controller D achieves QoE similar to that of the proposed policy but requires more than twice the encoded bitrate and higher glasses power consumption.}
    \label{fig:agentic_sim}
\end{figure}

As shown in Fig.~\ref{fig:agentic_sim}, the cellular-only baselines either stall at the O2I transition or cannot recover inside the RF dead zone during the indoor dwell. Switching bands only moves the connection to another blocked macro link. By contrast, all policies with access to the full cooperation mechanisms can continue streaming through the dead zone. The agentic policy switches bands before the link collapses, completes a make-before-break handover while the previous link remains available, and then delegates traffic to the CPE, whose local link remains usable. During the indoor dwell, it delegates the entire uplink whenever the shared CPE has spare capacity, reducing smartphone power consumption. Its predictive function also uses the thermal trajectory and expected scene complexity to reduce the encoded bitrate before the glasses approach the skin-contact temperature limit.

The results separate the contributions of cooperation, adaptivity, and agentic control. \emph{Cooperation} removes the stalls, and the capability-matched myopic controller independently selects low-band relaying and full CPE delegation when they are beneficial. \emph{Full-information adaptivity} is sufficient to achieve nearly the same instantaneous QoE as the proposed policy. \emph{Prediction and intent interpretation}, however, reduce the resources required to maintain that quality. The myopic controller uses an average encoded bitrate of 18.6~Mbps, compared with 8.8~Mbps for the proposed policy, and operates close to the thermal limit. After a 15-minute session, the glasses retain 32\% battery under the myopic controller and 58\% under the proposed policy. Over longer multi-trajectory episodes, the glasses under the myopic controller deplete their battery at about 28 mins, whereas the proposed policy continues streaming at the target quality at 36 mins and provides 22--47\% more useful quality-seconds. The ablation without the predictive governor exceeds the temperature limit.

\subsection{Wearable-Collaborative Blind-Spot Sensing and Warning}

\paragraph{A General Description}
The same device cluster can also support safety-aware sensing. As depicted in Fig.~\ref{fig:use_cases}, the creator rides an electric scooter, with the smartphone serving as the hub agent and a scooter-mounted CPE as an assisting device. While first-person streaming remains active, the cluster can monitor the rear blind spot for approaching cyclists or vehicles. The arm-mounted smartphone transmits a probing signal toward the rear, while the smartwatch or CPE receives the reflected echoes. This bistatic geometry can reduce self-interference relative to a monostatic device of similar size and provide a wider virtual sensing aperture. Because sensing has different requirements from streaming, the hub may switch to a safety-priority intent that limits the video enhancement layer, reserves resources for sensing tokens, and duplicates warning packets over both uplink paths.

Semantic-token aggregation reduces the amount of information exchanged among the devices. Instead of forwarding raw samples, each assisting device reports compact tokens that the hub fuses with observations from the glasses and edge-side map information. When sensing confidence decreases, the hub can recruit another device or reassign transmitter and receiver roles. The resulting output is an actionable warning, such as \emph{cyclist approaching from the rear blind spot}, together with confidence, direction, and urgency.

\paragraph{An Illustrative Scenario Study}

We evaluate this use case through a system-level simulation that follows the same ride but focuses on the sensing path. The smartwatch is the default bistatic receiver, and the scooter-mounted CPE serves as an auxiliary receiver. A cyclist approaches from the rear blind spot, followed by two successive disturbances. First, the rider's arm blocks the smartwatch echo path. Later, the CPE becomes temporarily unavailable.

The sensing receive chain of the smartwatch consumes non-negligible power relative to its battery capacity of approximately 1~Wh. Once the scooter-powered CPE provides reliable rear coverage, disabling the smartwatch sensing chain is therefore energy-efficient for an adaptive controller. The main difference among the evaluated policies is how they respond when the delegated CPE becomes unavailable.

We compare four policies: a no-cooperation baseline using the always-on smartwatch; a static baseline that delegates sensing to the CPE for the entire encounter without closed-loop recovery; a \emph{capability-matched reactive controller}; and the proposed agentic policy. The reactive controller uses the same confidence threshold for CPE recruitment and has access to the same sensing actions. However, without the micro-agent failure token, it detects CPE loss only after a report timeout and does not change its communication intent. The complete geometry, link budget, waveform-level detection model, and control logic are provided in the supplementary material \cite{WenSupplement2026}.

\begin{figure}[t]
    \centering
    \includegraphics[width=\linewidth]{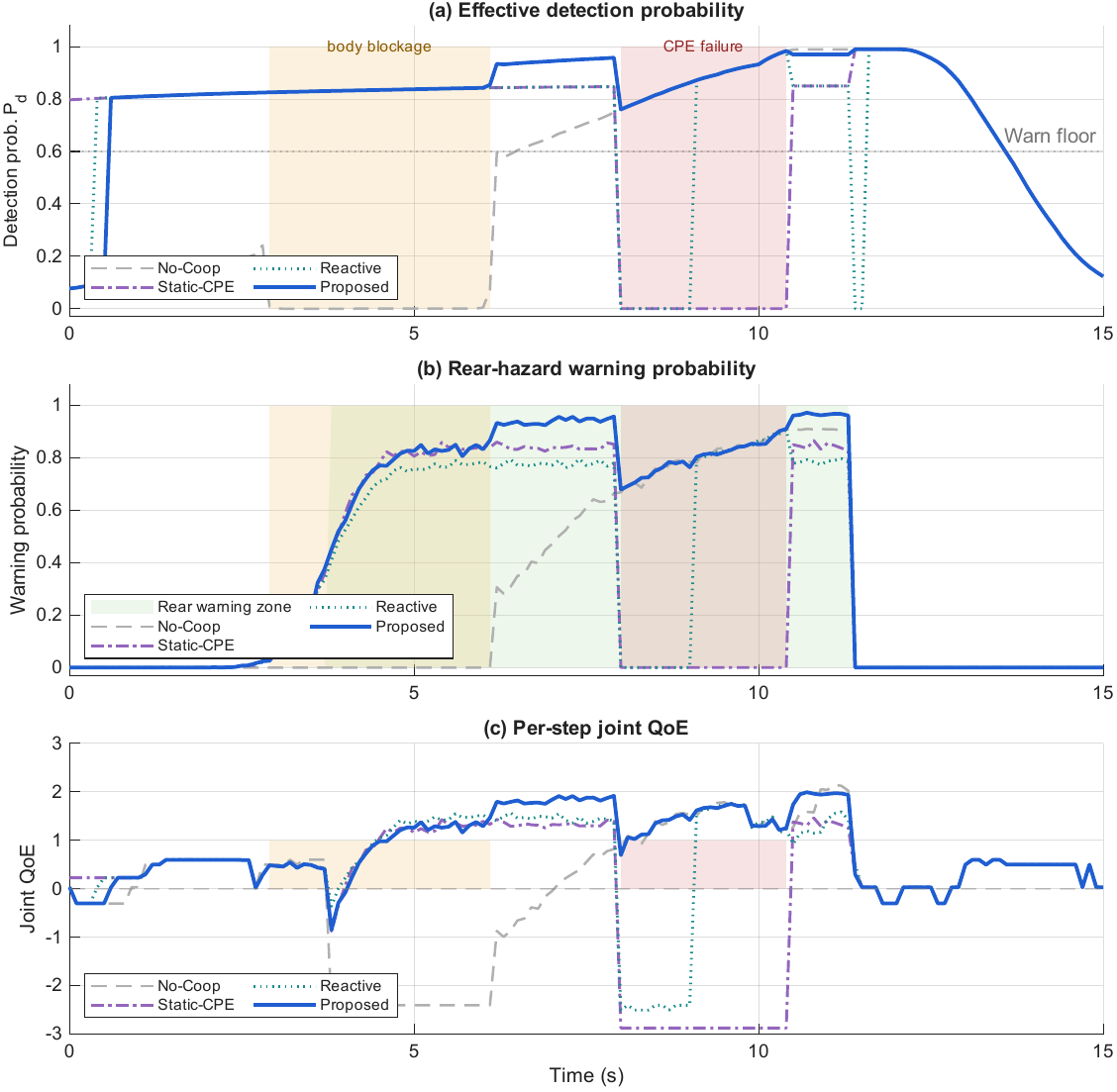}
    \caption{System-level evaluation in the wearable blind-spot sensing scenario with two successive disturbances: body blockage of the smartwatch echo path (left shaded band) and a transient scooter-CPE failure (right shaded band). The panels show (a) effective detection probability, (b) rear-hazard warning probability inside the warning zone, and (c) per-step joint QoE for a watch-only baseline (No-Coop), static delegation (Static-CPE), a capability-matched reactive controller, and the proposed agentic policy. Only the agentic policy maintains a continuous warning through both disturbances.}
    \label{fig:blindspot_sim}
\end{figure}

As shown in Fig.~\ref{fig:blindspot_sim}, the policies respond similarly to the initial body blockage but differ after the CPE failure. During body blockage, the mean smartwatch detection probability drops to near zero, and the no-cooperation baseline loses the rear-hazard warning. The cooperating policies instead use the CPE, whose detection probability remains near 0.84, and generate a warning within approximately 0.1~s after the target enters the warning zone. When the CPE subsequently fails, the static baseline remains without a warning until the target has passed. The reactive controller recovers only after the report timeout and smartwatch wake-up, resulting in a warning gap of more than one second. The agentic hub identifies the failure from the micro-agent token and immediately returns to smartwatch-assisted sensing. Because the safety intent keeps the smartwatch available as a redundant sensing path, the warning remains continuous.

The agentic policy maintains a reliable warning for 98.7\% of the interval in which the target is inside the warning zone, compared with 82.9\% for the reactive controller, 65.8\% for static delegation, and 56.6\% for no cooperation. It also achieves the highest joint QoE while consuming less wearable energy than the always-on-watch baseline. These results again separate the contributions of cooperation and agentic control. Cooperation restores sensing capability during blockage, while failure attribution and intent-dependent redundancy allow immediate recovery when an assisting device becomes unavailable.

These two examples illustrate a broader class of applications. Enterprise field service, construction-site safety, and immersive event broadcasting require similar decisions on assisting-device selection, communication or sensing roles, compute placement, and task-aware information exchange under device and network constraints.

\section{Standardization and Open Research Challenges}

Agentic UE-CoMIMO coordinates nearby devices, CPEs, and edge resources as a virtualized terminal and therefore requires support beyond physical-layer UE-CoMIMO. These requirements have different levels of maturity. Table~\ref{tab:std} distinguishes, for each interface, existing mechanisms, implementable extensions of current mechanisms, and genuinely new standardization requirements.

\begin{table*}[t]
\centering
\caption{Standardization needs of Agentic UE-CoMIMO: existing mechanisms, implementable extensions, and genuinely new requirements.}
\label{tab:std}
\footnotesize
\begin{tabular}{p{2.9cm}p{4.2cm}p{4.4cm}p{4.6cm}}
\toprule
Interface & Existing mechanisms & Implementable extension & Genuinely new requirement \\
\midrule
Device discovery
& 3GPP sidelink discovery; Wi-Fi peer discovery
& Capability exposure extended to relay, sensing, and computing capabilities, together with battery and thermal states
& Cross-trust-domain capability discovery with explicit user-consent support \\ \hline
Local control authority
& UE capability reporting; network-configured measurement and reporting frameworks
& Network-signaled parameter envelopes, such as allowed bands, power levels, and duty cycles, within which the hub can act autonomously
& A standardized policy-envelope contract that permits fast local decisions subject to network monitoring without per-decision approval \\ \hline
Third-party CPE participation
& Release-17 UE-to-network relaying \cite{Ganesan2021Sidelink}
& CPEs and APs acting as authorized relays or sensing receivers for registered users
& Authorization, accounting, incentive, and liability mechanisms for assisting devices not owned by the user \\ \hline
Relay modes
& L2/L3 sidelink relaying
& Control-plane support for L1 frequency-translation relays
& A unified switching abstraction across relaying, splitting, duplication, and token-exchange modes \\ \hline
Semantic-token exchange
& CSI and sensing feedback frameworks as the closest existing mechanisms
& Token metadata carried over existing QoS or assistance signaling
& A standardized token header, including task identity, uncertainty, freshness, trust, and resolution tier, together with exchange procedures, while leaving the payload implementation-specific \\
\bottomrule
\end{tabular}
\end{table*}

Two aspects deserve particular attention. The first is the division of \emph{control authority}. If every decision is made by the network, responses to blockage, mobility, or overheating may be too slow. If the hub operates without constraints, it may cause excessive duplication, interference, or unfair resource usage. A practical solution is the policy envelope introduced in Section~\ref{sec:architecture}. The network defines parameter ranges, resource budgets, and security policies, while the hub makes fast local decisions within these limits. Such decisions can be monitored by the network without requiring per-decision approval. 

The second aspect is \emph{token interoperability}. Standardizing the token header and exchange procedures described in Section~\ref{sec:control_and_semantic}.B, while leaving the payload representation implementation-specific, can support multi-vendor interoperability without constraining future encoder designs.

Trust and privacy affect all of these interfaces. Because personal devices, CPEs, and public infrastructure may jointly sense and communicate, first-person video, sensing tokens, and device-state reports can expose sensitive information. An untrusted Co-UE may also inject false tokens or manipulate sensing confidence. Relay authentication, secure token exchange, explicit user consent, and malicious-device detection are therefore required, and trust should be considered when recruiting assisting devices.

Agentic policies also require validation beyond simulation because cooperation depends on local geometry, body blockage, mobility, and hardware impairments. This calls for OTA-calibrated models, measurement-based evaluation, and multi-device experiments. The control design should also remain stable under partial observability by using conservative fallback actions when confidence is low and by limiting the rate of topology changes to avoid oscillation.

\section{Conclusion}
\label{sec:conclusion}

UE-CoMIMO has shown that multiple user-side devices can form a virtual antenna system to overcome the form-factor limitations of an individual terminal. This article extends this concept to Agentic UE-CoMIMO, in which a user-side control layer selects cooperating devices, assigns their roles, determines relay and traffic modes, places computation, and allocates token budgets according to user intent and device state. The two scenario studies further separate the contributions of cooperation, adaptivity, and agentic control. Cooperation provides the underlying capability, while adaptive control can achieve similar instantaneous performance. Prediction, intent interpretation, and feedback-driven replanning improve resource sustainability over longer sessions and maintain service continuity when an assisting device fails.

\bibliographystyle{IEEEtran}
\bibliography{References}






\end{document}